\documentclass[%
 reprint,
superscriptaddress,
 amsmath,amssymb,
 aps,
]{revtex4-2}

\usepackage[english]{babel}
\usepackage[letterpaper,top=2cm,bottom=2cm,left=3cm,right=3cm,marginparwidth=1.75cm]{geometry}
\usepackage{amsmath}
\usepackage{graphicx}
\usepackage[colorlinks=true,hyperfootnotes=true,breaklinks=true,citecolor=blue]{hyperref}
\usepackage{tabularray}
\usepackage{comment}

\usepackage[ruled]{algorithm2e}
\SetKwComment{Comment}{/* }{ */}

\begin{document}

\title{Universal low-depth two-unitary design of programmable photonic circuits}

\author{S.\,A.\,Fldzhyan}
\affiliation {Faculty of Physics, M.\,V. Lomonosov Moscow State University, Leninskie Gory 1, Moscow 119991, Russia}

\author{M.\,Yu.\,Saygin }
\email{saygin@physics.msu.ru}
\affiliation {Sber Quantum Technology Center, Kutuzovski prospect 32, Moscow 121170, Russia}
\affiliation {Faculty of Physics, M.\,V. Lomonosov Moscow State University, Leninskie Gory 1, Moscow 119991, Russia}

\author{S.\,S.\,Straupe }
\affiliation {Sber Quantum Technology Center, Kutuzovski prospect 32, Moscow 121170, Russia}
\affiliation {Faculty of Physics, M.\,V. Lomonosov Moscow State University, Leninskie Gory 1, Moscow 119991, Russia}

\begin{abstract}

The development of large-scale, programmable photonic circuits capable of performing generic matrix-vector multiplication is essential for both classical and quantum information processing. However, this goal is hindered by high losses, hardware errors, and difficulties in programmability. We propose an enhanced architecture for programmable photonic circuits that minimizes circuit depth and offers analytical programmability, properties that have not been simultaneously achieved in previous circuit designs. Our proposal exploits a previously overlooked representation of general nonunitary matrices as sums of two unitaries. Furthermore, similar to the traditional singular value decomposition-based circuits, the circuits in our unitary-sum-based architecture inherit the advantages of the constituent unitary circuits. Overall, our proposal provides a significantly improved solution for matrix-vector multiplication compared to the established approaches.

\end{abstract}

\maketitle

\section*{Introduction}

Programmable photonic circuits are increasingly used as an energy-efficient solution for classical information processing~\cite{PhotonicML,IterativeSolvers,PNN_survey} and play a crucial role in optical quantum  computing~\cite{CarolanUniversalLO,PsiQuantum2025,Xanadu2025,GraphProblems}. However, as analog devices, photonic circuits face challenges such as errors, losses, and limited programmability that hinder their scalability and prevent them from competing with traditional digital electronic methods.

The core mathematical operation that underpins the computational power of programmable multimode photonics is matrix-vector multiplication (MVM). This operation is vital for data transformation~\cite{PIC_communications,PIC_review,PIC_rf}, weight adjustment, and learning processes in photonic neural networks~\cite{PhotonicML}. Recent advancements in photonic architectures have largely focused on programmable interferometers that perform multiplications with unitary transfer matrices. Several unitary designs for programmable photonic circuits have been proposed and widely adopted by researchers. Notably, the designs by Reck \emph{et~al.~}\cite{ReckDesign} and Clements \emph{et~al.~}\cite{ClementsDesign} stand out, as they benefit from an analytical procedure that calculates the phase-shift values needed for a given target matrix, assuming the circuits are error-free. Additionally, methods for error correction in unitary programmable circuits have been suggested~\cite{HamerlyErrorCorrection}, along with original circuit designs that maintain universality even in the presence of high error rates~\cite{FldzhyanET,Robust,AsymptoticallyHamerly}.

However, many information processing tasks require MVM involving a broader range of matrices beyond the unitary group. This is common in classical photonic neural networks~\cite{PNN_survey}, iterative solvers~\cite{IterativeSolvers}, and quantum graph problem solvers~\cite{GraphProblems}. Two main approaches have been proposed to implement photonic multiplication by nonunitary matrices. The first is the well-established singular value decomposition (SVD) method, which uses a sequence of a programmable unitary multimode circuit, mode wise amplitude modulation, and another unitary circuit~\cite{SVD_book}.  The advantage of this approach is its straightforward programmability, inherited from the analytically programmable unitary circuits.

The second approach involves embedding the target nonunitary matrices into larger unitary matrices~\cite{TangLowDepth,LowDepth,AsymptoticallyHamerly}. These methods typically require shorter circuit depths compared to SVD-based circuits. For example, Tang \emph{et~al.~}\cite{TangLowDepth} recently proposed embedding target nonunitary matrices into unitary circuits of half the depth of the corresponding SVD-based circuits. More recently, we proposed an even more compact architecture that is compatible with planar integrated photonics~\cite{LowDepth}. However, both approaches sacrifice the ease of analytical programmability, making their practical use challenging due to the lack of effective programming algorithms and \emph{in~situ} training methods. This, in turn, can spoil all the computational advantage of photonic MVM.

In this work, we propose a programmable photonic circuit architecture that significantly improves both the SVD-based circuits~\cite{SVD_book} and our recent low-depth architecture~\cite{LowDepth}. The proposed architecture is based on representing the target nonunitary matrices as the sum of two unitary matrices.  Our approach synergetically combines the advantages of both the SVD-based and embedding-into-unitary methods. Specifically, the programmable circuits derived from our architecture are both low depth and analytically programmable.

\section{Low-depth circuit}

We are interested in generic MVM operation, in which the input vector of field amplitudes $\boldsymbol{a}^{(in)}=(a^{(in)}_1,\ldots,a^{(in)}_N)^{\text{T}}$ is multiplied by a target $M\times{}N$ complex-valued matrix $W$:
    \begin{equation}\label{eqn:mvm}
        \boldsymbol{a}^{(\text{out})}=W\boldsymbol{a}^{(\text{in})},
    \end{equation}
in the course of their propagation through a programmed multiport linear-optical circuit to obtain the result $\boldsymbol{a}^{(\text{out})}=(a^{(\text{out})}_1,\ldots,a^{(\text{out})}_M)^{\text{T}}$ at its output. In the following, we assume square matrices $W$ with $N=M$, which will not undermine the generality of our results. The general task of performing MVM by optics boils down to constructing a programmable multiport circuit that can implement a class of transfer matrices of interest $W$.

\subsection{SVD-based architecture}

\begin{figure}
    \centering
    \includegraphics[width=0.95\linewidth]{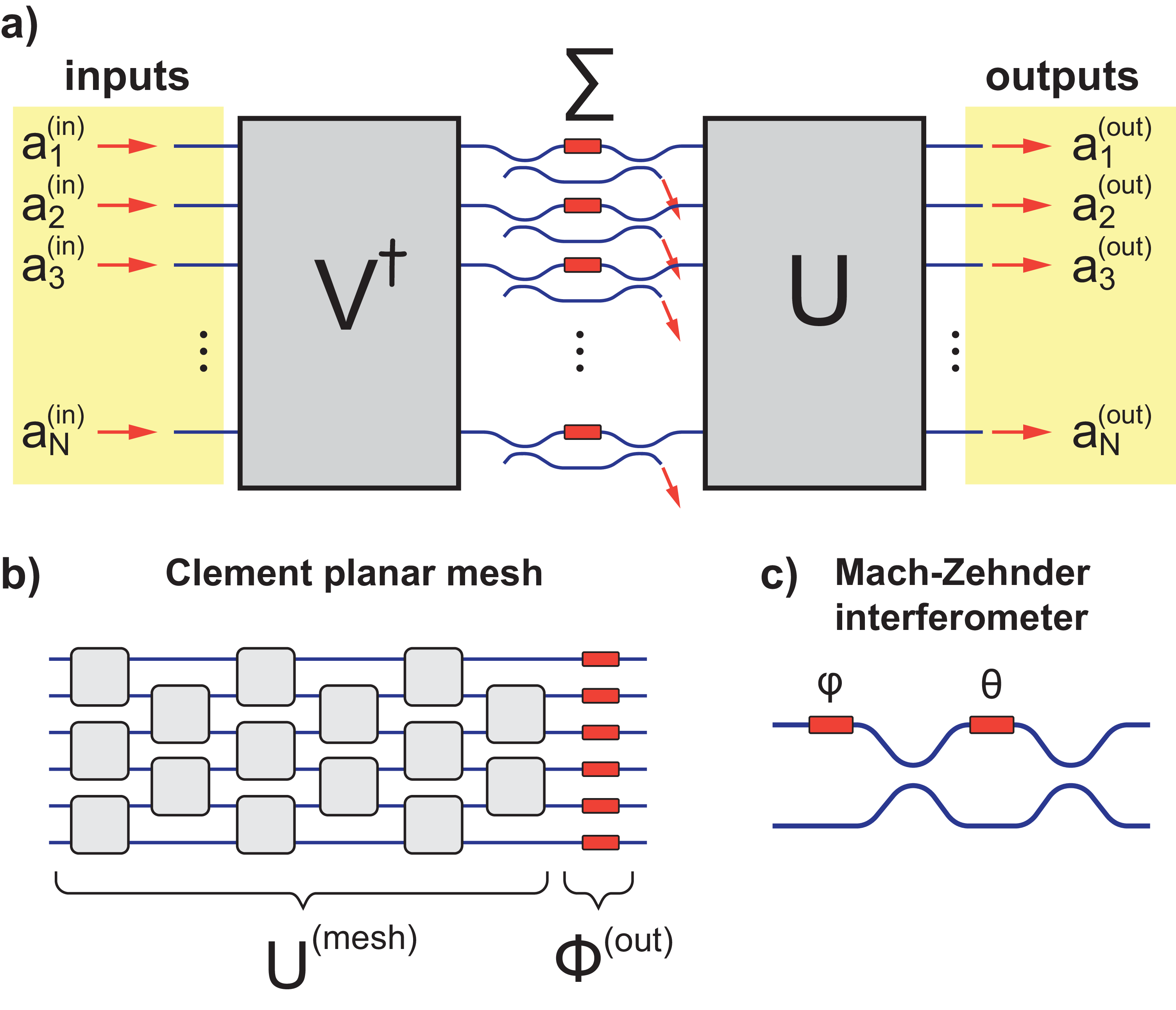}
    \caption{Traditional method of the universal MVM operations: (a) the SVD-based programmable circuit consisting of two programmable unitary circuits and programmable amplitude modulators connected in sequence, (b) the universal planar unitary mesh built from two-mode blocks $U^{(\text{mesh})}$ and a phases-shift layer $\Phi^{(\text{out})}$, and (c) example of two-mode block---Mach-Zehnder interferometer.}
    \label{fig:svd}
\end{figure}

In this work, we are interested in the universal MVM---the most general form of MVM, in which the circuit is required to implement arbitrary complex-valued matrices $W$, which can be parametrized by $2N^2$ real parameters. To this date, the most often exploited approach to accomplish this goal has been using the SVD-based circuits, depicted in Fig.~\ref{fig:svd}(a). The SVD represents any nonunitary complex-valued matrix $W$ in the form of the product of three matrices:
    \begin{equation}\label{eqn:svd}
        W=U\Sigma{}V^{\dagger},
    \end{equation}
each of which is implemented by a distinct photonic unit, as shown in Fig.~\ref{fig:svd}(a). In Eq.~\eqref{eqn:svd}, $U$ and $V$ are unitary matrices
and $\Sigma=\text{diag}(\sigma_1,\ldots,\sigma_N)$ is the diagonal matrix consisting of the real singular values $\sigma_j$ of the matrix $W$. The matrix $\Sigma$ is physically realized using programmable amplitude damping. Strictly speaking, for arbitrary matrices $W$, the singular values $\sigma_j$ are unbounded; however, the linear optical implementation of the SVD imposes the constraint $|\sigma_j|\le{}1$. Fortunately, this does not diminish the applicability of linear optical implementations, since the target matrices $W$ can be rescaled to meet the constraint. The multiplication by arbitrary unitary matrices $U$ and $V$ can be realized using several interferometer designs. In integrated photonics, the conventional designs are universal planar meshes $U^{(\text{mesh})}$ of two-mode programmable blocks and a phase-shift layer $\Phi^{(\text{out})}$, as shown in Fig.~\ref{fig:svd}(b). One particular example of the mesh is the Clements design~\cite{ClementsDesign} consisting of Mach-Zehnder interferometers, shown in Fig.~\ref{fig:svd}(c).

An important characteristic of programmable photonic circuits is their depth, which we measure by the number of layers of programmable phase-shift elements. This is justified because the programmable elements are typically the main contributor to optical losses and geometric size of the circuits, and the number of static BS components is proportional to this quantity. For the SVD-based circuit depicted in Fig.~\ref{fig:svd}(a) the depth is the sum of the depths of sequential unitaries $V^{\dagger}$ and $U$ and amplitude modulators $\Sigma$; one has $\mathfrak{D}_{\text{SVD}}=\mathfrak{D}_{V}+\mathfrak{D}_{U}+\mathfrak{D}_{\Sigma}$. To obtain $\mathfrak{D}_{\text{SVD}}$, we recall that the implementations of the most compact universal $N$-mode unitaries require $N+1$ phase-shift layers~\cite{Robust,BellCompactifying}. Also notice that in Eq.~\eqref{eqn:svd} one unitary circuit can be nonuniversal, as it can lack one phase-shift layer---either output of circuit $V^{\dagger}$ or input of circuit $U$. Therefore, one arrives at
    \begin{equation}\label{eqn:svd_depth}
        \mathfrak{D}_{\text{SVD}}=2N+2.
    \end{equation}

\subsection{Proposed two-unitary architecture}

The circuit we propose in this work is based on the simple observation that the product \eqref{eqn:svd} can be rewritten as a sum of two unitary matrices. To show this, we rewrite the diagonal matrix $\Sigma$ as follows:
    \begin{equation}\label{eqn:sigma_sum}
        \Sigma=\frac{1}{2}(D+D^{\ast}),
    \end{equation}
where $D=\text{diag}(e^{i\psi_1},\ldots,e^{i\psi_N})$ is the diagonal unitary matrix with $e^{i\psi_j}=\sigma_j+i\sqrt{1-\sigma_j^2}$ ($j=\overline{1,N}$). Then, substituting Eq.~\eqref{eqn:sigma_sum} into Eq.~\eqref{eqn:svd}, we come to the expression
    \begin{equation}\label{eqn:sum_matrix}
        W=\frac{1}{2}(U{}D{}V^{\dagger}+U{}D^{\ast}V^{\dagger})=\frac{1}{2}(U^{(1)}+U^{(2)}).
    \end{equation}
Notice that in the ideal case when the unitary circuits can be programmed to arbitrary unitary matrices, programming is fully analytical and requires one SVD of $W$ and two matrix multiplications to produce $U^{(1)}$ and $U^{(2)}$ with complexity $O(N^3)$. We provide the pseudocode for the programming algorithm in Appendix~\ref{app:alg}.

\begin{figure*}
    \centering
    \includegraphics[width=0.7\linewidth]{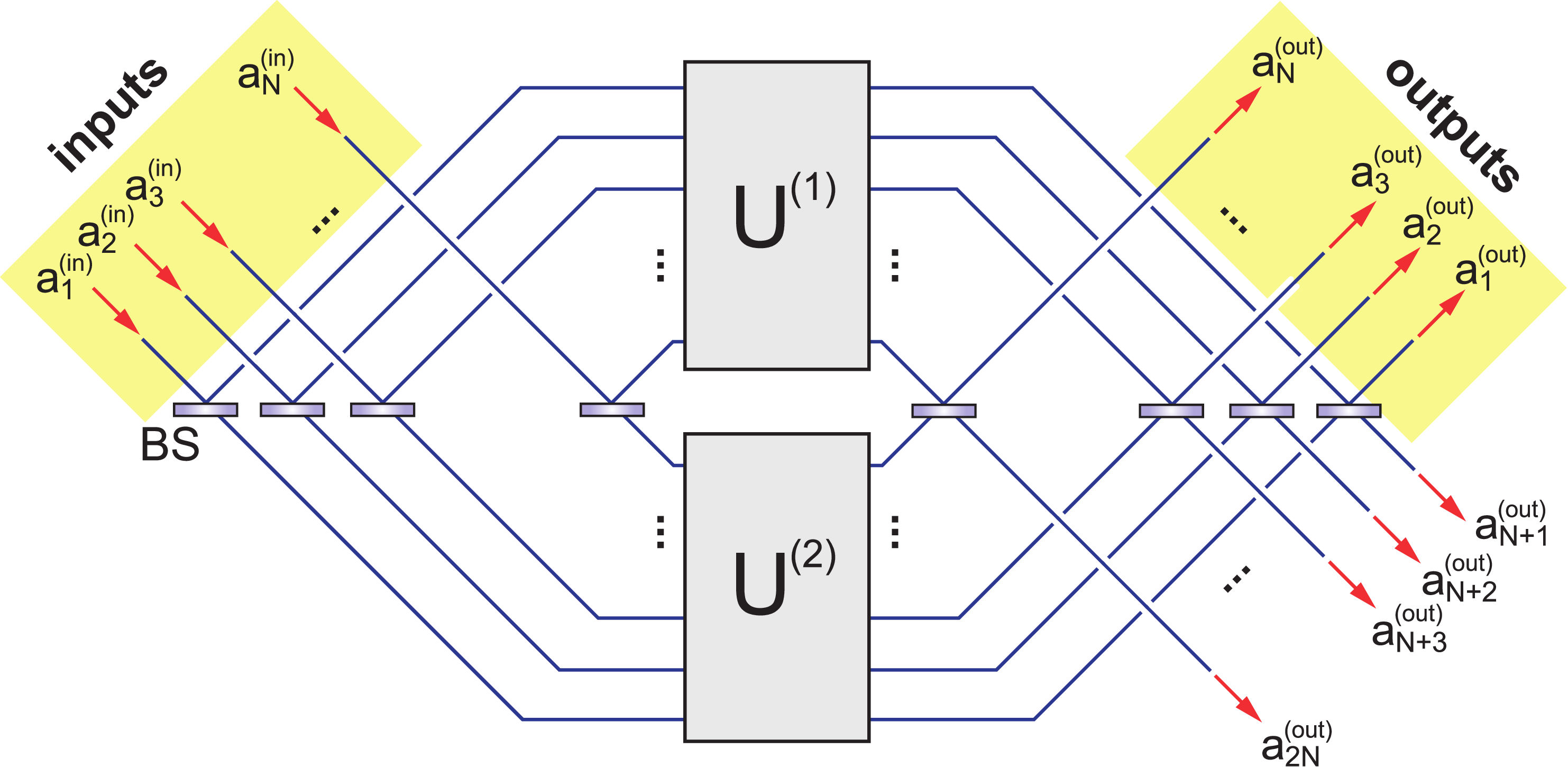}
    \caption{Proposal: The programmable circuit implementing the sum of two unitaries $U^{(1)}$ and  $U^{(2)}$ proposed to perform multiplication by $N\times{}N$ nonunitary matrices. Here, the BSs before and after unitary circuits are balanced.}
    \label{fig:unitary_sum}
\end{figure*}

The optical scheme performing MVM according to Eq.~\eqref{eqn:sum_matrix} is shown in Fig.~\ref{fig:unitary_sum}. The main part of the scheme is formed by two $N$-mode unitary circuits, $U^{(1)}=U{}D{}V^{\dagger}$ and $U^{(2)}=U{}D^{\ast}V^{\dagger}$, acting in parallel rather than in sequence, as in the SVD-based circuits shown in Fig.~\ref{fig:svd}. The MVM multiplication proceeds by first splitting the field vector to be multiplied $\boldsymbol{a}^{(\text{in})}$ on a set of $N$ balanced beam splitters (BSs), so that its each component is equally distributed among the unitary circuits. Without loss of generality, we take the beam-splitter transfer matrix as 
    \begin{equation}\label{eqn:bs}
        U_{\text{BS}}=\left(
        \begin{array}{cc}
            \sqrt{R} & \sqrt{1-R} \\
            \sqrt{1-R} & -\sqrt{R}
        \end{array}
        \right),
    \end{equation}
where $R$ is the power reflectivity of the BS that should ideally be balanced (i.e., $R=R_0=1/2$). In the following, we write $R=\cos^2(\pi/4+\alpha)$, where $\alpha$ quantifies 
the deviation of the BS from the balanced operation. After multiplication by the unitary circuits the amplitudes are combined on yet another set of balanced beam-splitter to produce $2N$ amplitudes $a^{(out)}_1,\ldots,a^{(\text{out})}_{2N}$. Then, the MVM result Eq.~\eqref{eqn:mvm}  is carried by the $N$ modes $a^{(\text{out})}_1,\ldots,a^{(\text{out})}_{N}$.

The circuit depth $\mathfrak{D}_{2U}$ of the two-unitary circuits, defined as the number of programmable phase-shift layers, is 
    \begin{equation}\label{eqn:twoU_depth}
        \mathfrak{D}_{2U}=N+1=\frac{1}{2}\mathfrak{D}_\text{SVD},
    \end{equation}
i.e., half the depth of the SVD-based realization \eqref{eqn:svd_depth}. Note that this depth metric does not capture additional loss or loss imbalance introduced by waveguide crossers that may be required in a single-layer planar implementation [Fig.~\ref{fig:unitary_sum}, two-dimensional layer]. Ultra low-loss crossers have been demonstrated~\cite{CrossersReview,Crossers2,CrossersTheory}; moreover, multilayer integrated platforms~\cite{TwoLayer,TriLayer} allow the two programmable unitaries $U^{(1)}$ and $U^{(2)}$ to be placed on separate layers and coupled interlayer, eliminating crossers altogether. In such multilayer implementations, the footprint and loss are then governed by the reduced depth relation above---i.e., Eq.~\eqref{eqn:twoU_depth} relative to Eq.~\eqref{eqn:svd_depth}---implying a twofold reduction in optical depth and size. Furthermore, the number of balanced beam splitters is the same in both Figs.~\ref{fig:svd} and \ref{fig:unitary_sum}. This means the depth reduction comes at no cost in additional beam splitters.

Therefore, the circuit that exploits two parallel programmable unitaries has the obvious advantage of having half the depth of its SVD-based counterpart, resulting in proportionally lower losses. It should also be noted that reducing the circuit depth may lead to decreased susceptibility to errors, such as unbalanced losses in the constituent elements, which accumulate with depth. Importantly, while being low depth, our circuit architecture retains the advantage of being easily reprogrammed according to a given target matrix without resorting to the computationally intensive numerical optimization inherent to other low depth circuits~\cite{LowDepth,TangLowDepth}.

\subsubsection{Tolerance to hardware errors}

Taking into account various error-tolerant designs for universal unitaries, such as circuits~\cite{FldzhyanET,Robust,AsymptoticallyHamerly}, the nonunitary circuit studied here can likewise be made error tolerant. The circuit’s error tolerance is achieved through static correction, not through dynamic error correction. This requires correcting errors due to static phase-shift imbalances in the paths between the BSs and the unitaries, as well as static imbalances in the BS splitting ratios. The former phase errors can always be corrected using the universal unitary circuits themselves. To show this, consider a target nonunitary matrix $W$ defined by the unitary matrices $U^{(1)}$ and $U^{(2)}$ in the decomposition Eq.~\eqref{eqn:sum_matrix}, which should be set in the unitary circuits provided no errors are present. Then, suppose phase errors are introduced before and after the unitary circuits, which are described by diagonal matrices $\Psi^{(j)}$ and $\Phi^{(j)}$ ($j=1,2$). Obviously, programming the unitaries into the matrices $\tilde{U}^{(j)}=\Phi^{(j)\ast}U^{(j)}\Psi^{(j)\ast}$, which is always possible due to the universality of the circuit, will correct these errors.

In contrary, the errors BSs that come in the form of deviations from balanced operation in Eq.~\eqref{eqn:bs} are not generally corrected by the unitary circuits themselves. However, this can be achieved by replacing each of the $2N$ BSs with a programmable Mach-Zehnder interferometer, introducing a linear overhead with respect to $N$, which is relatively small compared to the quadratic overhead from the programmable elements in the unitaries. However, in our study, we focus on the impact of hardware errors while assuming that these BSs remain static, but the transfer matrices $U^{(1)}$ and $U^{(2)}$ are unaffected by errors. We do not assume a specific error-tolerant photonic architecture for the unitary circuits; instead, we treat them as parametrized matrices. Specifically, we parametrize the unitaries  $U^{(1)}$ and $U^{(2)}$ by $2N^2$ parameters of the nonunitary matrix $K=\frac{U^{(1)}+U^{(2)}}{2}$ with singular values not exceeding $1$.  It should be noted that if no hardware errors are present, $K=W$; however, in the general erroneous, case $K\neq W$.   Appendix~\ref{app:Err case} details this parametrization for the erroneous case.

Thus, the only faulty elements are the BSs~\eqref{eqn:bs}, with the values of $\alpha$ representing the corresponding errors. Two limiting cases of error models were considered: (1) correlated errors, where all BSs have the same error $\alpha$, and (2) independent random errors, where the error $\alpha_j$ for each BS follows a normal distribution $\sim\exp(-\alpha_j^2/2\sigma^2)$, with $\sigma$ representing the degree of randomness in the errors.

\begin{figure*} 
    \centering 
    \includegraphics[width=0.95\linewidth]{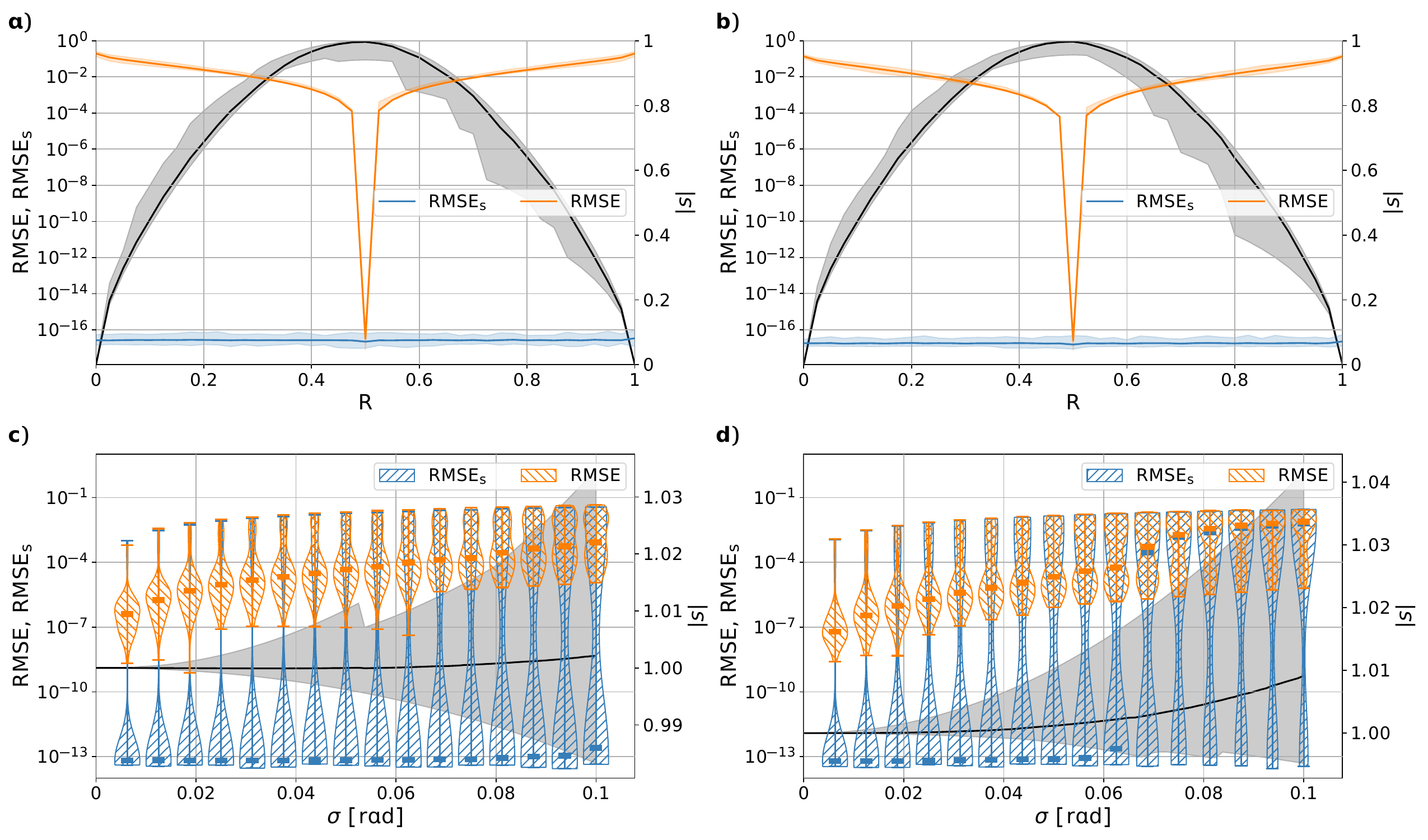}     
    \caption{\textbf{Effect of beam-splitter errors on the quality of photonic MVM operations}: The distributions of $\mathrm{RMSE}$ and $\mathrm{RMSE_s}$  (drawn in orange and blue) are shown (1) correlated errors as a function of the BS power reflectivity $R$ at $N=10$ [panel (a)] and $N=20$ [panel (b)], and (2) independently sampled random errors as a function of the distribution width $\sigma$ at $N=10$ [panel (c)] and $N=20$ [panel (d)].  Each $\mathrm{RMSE}$, $\mathrm{RMSE_s}$ distribution is based on $100$ randomly sampled target matrices (left axes). For the distributions corresponding to independently sampled random errors, the minimum, maximum, and median bars are plotted.  The distributions of $|s|$ corresponding to $\mathrm{RMSE_s}$ are also plotted in the figures (right axes): The black curves are the mean values, and the shaded regions are specified by the minimum and maximum values. The kinks in the shaded regions are an intrinsic feature of the underlying statistical randomness.} \label{fig:errors} 
\end{figure*}
To assess how accurately the actual circuit implements the MVM under BS errors, we compare the target matrices $W^{(0)}$ with the corresponding transfer matrices $W$ realized by the circuit. For this, we use the root mean square error (RMSE): 
    \begin{equation}\label{eqn:nse} 
        \text{RMSE}(W^{(0)},W)=\frac{1}{N}\sqrt{\sum_{i,j=1}^N|W^{(0)}_{ij}-W_{ij}|^2}, 
    \end{equation} 
where $W_{ij}$ and $W_{ij}^{(0)}$ are the elements of the matrices. Additionally, we use a more relaxed measure of matrix closeness, allowing $W^{(0)}$ and $W$ to coincide up to a global scaling factor $s$. The corresponding measure of closeness is $\mathrm{RMSE_s}(W^{(0)},W)=\mathrm{RMSE}(sW^{(0)},W)/|s|$~\cite{LowDepth}. Smaller values of $|s|$ are associated with larger losses, so they are penalized by $|s|$ in the denominator in $\mathrm{RMSE_s}$. To determine the parameters of the unitary circuits that implement a target nonunitary matrix $W^{(0)}$ with the lowest possible $\mathrm{RMSE}$ or $\mathrm{RMSE_s}$, we employed the Limited-memory Broyden-Fletcher-Goldfarb-Shanno (L-BFGS) optimization algorithm~\cite{2020SciPy-NMeth}. This algorithm scales well to high-dimensional parameter spaces with limited memory and minimal hyper-parameter tuning (see, e.g., Refs.~\cite{FldzhyanET,CompactBellState}).

We used SVD Eq.~\eqref{eqn:svd} to generate a target matrix $W^{(0)}$, wherein the unitary complex-valued matrices $U$ and $V$ were drawn from the Haar random distribution using the method based on the QR decomposition of random matrices from the Ginibre ensemble~\cite{RandomMatrices}. The diagonal matrix $\Sigma$ was filled with independently generated values from a uniform distribution in the $[0,1]$ range, which are then rescaled in such a way that the maximum value of $\sigma_j$ is equal to $1$.

Fig.~\ref{fig:errors} shows the results of the simulation. In Figs.~\ref{fig:errors}(a) and \ref{fig:errors}(b) the $\mathrm{RMSE}$ and $\mathrm{RMSE_s}$ distributions are plotted as a function of the power reflectivity for the circuit sizes of $N=10$ and $N=20$, respectively. As seen in the figures, even a small deviation of $R$ from its balanced value of $1/2$ degrades the quality of the matrix operation (the perfect  $\mathrm{RMSE}$ plateau at the level of $10^{-16}$ is set by finite machine precision). However, when the transfer matrix is allowed to scale by the global factor $s$, the correlated errors can be perfectly corrected by the unitary circuits and measure the quality of the MVM by $\mathrm{RMSE_s}$. Notice that the absolute value of the scaling factor required to correct correlated errors monotonically decreases with increasing value of error.

Figures~\ref{fig:errors}(c) and \ref{fig:errors}(d) suggest different behavior of the MVM quality, as quantified by both unscaled and scaled measures. However, the implementation of scaled matrices, quantified by $\mathrm{RMSE_s}$, is much better than that of unscaled matrices, quantified by $\mathrm{RMSE}$. It should be noted that this correction is attained by scaling factors that differ from $1$ by small values.

\section{Conclusion}


We have introduced a universal two-unitary architecture for programmable photonic matrix-vector multiplication that reduces circuit depth by a factor of 2 relative to SVD-based realizations, while retaining analytical programmability and full expressivity. A key feature of our approach is its architecture agnosticism: The two parallel unitaries can be instantiated by any universal unitary mesh or interferometer platform, including the canonical triangular and rectangular layouts of Reck \emph{et al.}~\cite{ReckDesign} and Clements \emph{et al.}~\cite{ClementsDesign}, as well as alternative integrated~\cite{FldzhyanET,HamerlyErrorCorrection} or free-space implementations. This decouples our contribution from a particular unitary design and makes it directly compatible with ongoing advances in mesh topology, calibration, and fabrication.

By halving the unitary circuit depth and confining additional components to a small, linear in $N$ overhead, the proposed scheme lowers accumulated loss and relaxes tolerance requirements in large-scale systems. These properties are advantageous across classical and quantum use cases. In classical analog computing and photonic neural networks, the combination of shallow depth and closed-form programming reduces both power and calibration burden. In quantum photonics, large interferometers underpin models such as boson sampling and Gaussian boson sampling, where optical depth and loss critically determine scalability. Our architecture serves as a drop-in alternative that preserves universality while improving depth and loss profiles.

Looking ahead, we anticipate that the two-unitary framework, together with standard techniques for mesh calibration, active stabilization, and (where available) multilayer photonic integration, will enable deeper circuits with lower error budgets.

\section{Acknowledgment}

S.A.F. is grateful to the Russian Foundation for the Advancement of Theoretical Physics and Mathematics (BASIS) (Project №23-2-10-15-1).

\appendix
\section{Algorithm}\label{app:alg}
\begin{algorithm}
\caption{Two-unitary algorithm pseudocode.}\label{alg:two}
\KwData{$W$\Comment*[r]{Target matrix}}

$\textbf{Begin:}$

\flushleft $U,\ \Sigma,\ V_{hc} := \text{svd}\left(W\right)$ \Comment*[l]{SVD algorithm with preferred software package}

$\Sigma := \Sigma/\max{\Sigma}$\;

$U_1 := U\cdot\left(\Sigma+i\sqrt{I-\Sigma^2}\right)\cdot V_{hc}$\;

$U_2 := U\cdot\left(\Sigma-i\sqrt{I-\Sigma^2}\right)\cdot V_{hc}$\;

\KwResult{$U_1, U_2$\Comment*[r]{Decomposition}}
\end{algorithm}
The pseudocode of the decomposition algorithm is presented in Algorithm~\ref{alg:two}. There, $\cdot$ designates matrix multiplication, and square root is taken elementwise.

Strictly speaking, any diagonal matrix $B_{ij}=\delta_{ij}(-1)^{b_i}$ with arbitrary bit string $b$ can be inserted to get a valid decomposition:
\begin{equation}
    U^{(1,2)}=U\left(\Sigma\pm i B \sqrt{I-\Sigma^2}\right) V^\dagger.
\end{equation}

\section{Erroneous case}\label{app:Err case}
Consider beam splitters with errors, represented by diagonal matrices of sines and cosines $S^{(i)},C^{(i)}$:
\begin{widetext}
\begin{multline}
    \begin{pmatrix}C^{(2)} & S^{(2)}\\ S^{(2)} & -C^{(2)}\end{pmatrix}\begin{pmatrix}U^{(1)} & 0\\ 0 & U^{(2)}\end{pmatrix}\begin{pmatrix}C^{(1)} & S^{(1)}\\ S^{(1)} & -C^{(1)}\end{pmatrix}=\\
    =\begin{pmatrix}C^{(2)}U^{(1)}C^{(1)}+S^{(2)}U^{(2)}S^{(1)} & C^{(2)}U^{(2)}S^{(1)}-S^{(2)}U^{(1)}C^{(1)}\\ S^{(2)}U^{(1)}C^{(1)}-C^{(2)}U^{(2)}S^{(1)} & C^{(2)}U^{(2)}C^{(1)}+S^{(2)}U^{(1)}S^{(1)}\end{pmatrix}.
\end{multline}
\end{widetext}
The transformation for the target $N$ modes can be expressed as
\begin{equation}\label{app:eq:err1}
    W = C^{(2)}U^{(1)}C^{(1)}+S^{(2)}U^{(2)}S^{(1)}.
\end{equation}
To streamline calculations, we avoid parametrizing unitary matrices using two sets of $N^2$ real parameters. Instead, we directly work with the $2N^2$ parameters of complex matrix coefficients. To achieve this, we introduce complex matrix $K$:
\begin{equation}
    K = \frac{U^{(1)}+U^{(2)}}{2}.
\end{equation}
Interestingly, when introducing an arbitrary matrix $K$ with singular values no greater than $1$, there exists matrix $J$ such that $K \pm iJ$ are unitary. It can be shown that given the singular value decomposition of $K$,
\begin{equation}
    K=R\Lambda Q\Rightarrow J = R\sqrt{I-\Lambda^2}Q
\end{equation}
is a solution. To ensure that the singular values of $K$ remain no greater than $1$, we modify the SVD decomposition as follows:
\begin{equation}
    \Lambda \rightarrow \Lambda/(1+\max\Lambda).
\end{equation}
This adjustment guarantees that all singular values are bounded by $1$.

Thus, the matrix $K$ alone suffices to parametrize our unitaries. Rewriting Eq.~\eqref{app:eq:err1} in terms of the new matrices and the error parameters $\alpha$, we obtain
\begin{equation}
    W_{ij}=K_{ij}\cos(\alpha^{(2)}_i-\alpha^{(1)}_j)-iJ_{ij}\sin(\alpha^{(2)}_i+\alpha^{(1)}_j),
\end{equation}
which can then be used with $W^{(0)}$ to compute $\mathrm{RMSE}$ and $\mathrm{RMSE_s}$.

\bibliography{biblio}

\end{document}